\documentclass[amsmath,amssymb, aps, prb, twocolumn, notitlepage]{revtex4-2}
\usepackage{graphicx}
\usepackage{dcolumn}
\usepackage{bm}
\usepackage[usenames]{color}
\usepackage{slashed}
\usepackage[utf8]{inputenc}
\usepackage{amsfonts}
\usepackage{amsmath}
\usepackage{amssymb}
\usepackage{hyperref}
\usepackage{latexsym}
\usepackage{color}
\usepackage{mhchem}
\usepackage{braket}
\usepackage{diagbox}
\usepackage{booktabs}
\usepackage{MnSymbol}
\usepackage{braket}
\usepackage{float}
\makeatletter
\let\newfloat\newfloat@ltx
\usepackage{algorithm}
\usepackage{algorithmicx}
\usepackage{algpseudocode}

\newcommand{\printfnsymbol}[1]{%
  \textsuperscript{\@fnsymbol{#1}}%
}

\begin{document}
\title{Many-electron characterizations of higher-charge superconductors}

\author{Zi-Hao Dong}
\affiliation{International Center for Quantum Materials, School of Physics, Peking University, Beijing, 100871, China}
\author{Yi Zhang}
\email{frankzhangyi@pku.edu.cn}
\affiliation{International Center for Quantum Materials, School of Physics, Peking University, Beijing, 100871, China}

\date{\today}

\begin{abstract}
The theoretical understanding of conventional superconductivity as the phonon-assisted formation and condensation of two-electron Cooper pairs is a significant triumph in condensed matter physics. Here, we propose many-electron characterizations of higher-charge superconductivity with Cooper pairs consisting of more than two electrons, by implementing translation symmetrization on parent pair-density-wave-ordered states. In particular, we demonstrate many-electron constructions with vanishing charge-2$e$ sectors, but with sharp signatures in charge-$4e$ or charge-$6e$ expectation values instead. Such characterizations are consistent with previous phenomenology of vestigial order and Ginzburg-Landau theory. Furthermore, we demonstrate that at the microscopic level, momentum conservation alone may be vital and sufficient for the robust emergence of higher-charge superconductivity. Our study thus offers a novel, general, and microscopic route to understand and characterize higher-charge superconductivity, including nontrivial experimental signatures such as fractional magnetic flux and period in interferometry, as well as localized Cooper pairs at lattice topological defects.
\end{abstract}

\maketitle

\emph{Introduction.}--- Superconductors naturally exhibit zero electrical resistance and the Meissner effect below a critical temperature \cite{onnes1911resistance, meissner1933neuer, van2010discovery}. Discovered by Onnes in 1911, the origin of superconductivity remained a microscopic mystery, despite the phenomenological success of its Ginzburg-Landau theory \cite{ginzburg2009theory}, until the BCS theory was proposed in 1957, which explains the dissipation-less current through the formation and condensation of Cooper pairs - two electrons bound together via lattice-mediated attractive interactions \cite{cooper1956bound, bardeen1957theory}. Such Cooper pairs, when generalized to finite momenta, give rise to pair density waves (PDWs), whose superconducting order parameter breaks translation symmetries and varies periodically in space \cite{fulde1964superconductivity, larkin1965nonuniform, agterberg2008dislocations, berg2009striped, lee2014amperean, fradkin2015colloquium, agterberg2020physics, wu2023pair, yerin2023multiple, ticea2024pair}. Signatures of such exotic phases have been established in recent experiments \cite{hamidian2016detection, liu2021discovery, chen2021roton, zhao2021cascade, gu2023detection}, including those involving the layered Kagomé material CsV$_3$Sb$_5$ \cite{chen2021roton, zhao2021cascade}.

Higher-charge superconductors can emerge as vestigial orders of an incommensurate PDW, which may be (partially) melted by disorder, thermal fluctuations, quantum fluctuations, or quasi-crystal structures \cite{agterberg2008dislocations, berg2009charge, agterberg2011conventional, jian2021charge, nie2014quenched, liu2024nematic}. Subsequently, the broken continuous translation symmetry is restored, and only the $U(1)$ gauge symmetry (charge conservation) breaking is retained. In a Ginzburg–Landau theory framework, these exotic superconducting phases reflect higher orders of the PDW order parameter fields, characterized by multiplets of Cooper pairs, such as with charges of $4e$ or $6e$ \cite{agterberg2008dislocations, berg2009charge, herland2010phase, agterberg2011conventional, jian2021charge, fernandes2021charge, han2022understanding, varma2023extended, zhang2024higgs, lin2025theory}. The corresponding fractional magnetic flux quanta, aside from alternatives such as multi-component or time-reversal-symmetry-breaking superconductors \cite{geshkenbein1987vortices, sigrist1989low, kee2000half, rampp2022integer}, have been potentially observed in quantum-interference experiments \cite{ge2024charge}. While theoretical mechanisms have been explored via the Hubbard models \cite{jiang2017charge, gnezdilov2022solvable, soldini2024charge, LI20242328}, vestigial fluctuations \cite{liu2023charge,volovik2024fermionic, wu2024d, song2025phase}, and geometric frustration \cite{zhou2022chern, pan2024frustrated, zeng2024high}, a comprehensive, microscopic many-electron characterization remains lacking, limiting our description and understanding of these fascinating quantum phenomena and correlated systems.

Candidate wavefunctions play a crucial role in understanding exotic quantum matters, often bypassing the need for exact ground-state solutions of intractable microscopic Hamiltonians. A landmark example is the Laughlin wavefunctions, which successfully characterize fractional quantum Hall liquids and encode the fractional statistics of topological order \cite{laughlin1983anomalous, arovas1984fractional, jain1989composite}, profoundly guiding subsequent quantum-transport and shot-noise experiments \cite{tsui1982two,saminadayar1997observation}. In the realm of quantum spin liquids, Gutzwiller-projected wavefunctions derived from slave-particle approaches offer a powerful variational ansatz \cite{gutzwiller1963effect, wen1991mean, lee2006doping, zhou2017quantum}, supporting an effective field theory for universal behaviors such as long-range entanglement and emergent gauge structures, and successfully predicting the gapless spinon Fermi surfaces later observed in organic materials \cite{yamashita2008thermodynamic,yamashita2010highly}. Overall, these wavefunctions serve not only as trial states for variational studies \cite{yokoyama1987variational1, yokoyama1987variational2, gros1989physics, paramekanti2001projected, himeda2002stripe, motrunich2005variational, edegger2007gutzwiller, iqbal2011projected, iqbal2013gapless} but also as conceptual bridges linking microscopic models to experimental observables of emergent quantum phases. Furthermore, the restoration of translation symmetry via projection is a well-established operation for recovering well-defined quantum numbers in variational studies across quantum chemistry \cite{jimenez2012projected}, nuclear physics \cite{nikvsic2006beyond, sogo2009critical}, and condensed matter physics \cite{mizusaki2004quantum, heitmann2019combined}.

In this work, we propose and study candidate many-electron quantum states of higher-charge superconductors, based on translation-symmetrizing parent PDW states. For the simplest scenario of a charge-$4e$ superconducting state rooted from a stripe PDW in two dimensions (2D), the resulting states display vanishing single-Cooper-pair sectors, yet finite expectation values over certain charge-$4e$ Cooper pairs, e.g., $\langle c^\dagger_{\mathbf{k},\uparrow}c^\dagger_{-\mathbf{k+Q},\downarrow}c^\dagger_{\mathbf{k'},\uparrow}c^\dagger_{\mathbf{-k'-Q},\downarrow} \rangle$, a composite of two Cooper pairs with momentum $\pm \mathbf{Q}$, respectively. Generalizations to even higher-charge states, such as charge-$6e$, and to higher dimensions, are also straightforward. The formalism is consistent with the physical picture of vortex binding between the PDW order parameter fields in previous vestigial order and classical Ginzburg-Landau theory studies. Moreover, while point-group symmetries often play a vital role in stabilizing the coexisting parent PDW states, we demonstrate that the underlying kinematics for the subsequent emergence of higher-charge superconductivity is fundamentally driven by momentum conservation. We further elaborate on this general mechanism in a perturbation theory.

Furthermore, we examine higher-charge superconductors with respect to an interferometry geometry and observe apparent periodicity in the threading magnetic fluxes associated with the proposed higher-charge Cooper pairs, which differs significantly from that of the parent charge-$2e$ PDW. Interestingly, we find that such higher-charge superconductors interact nontrivially with certain topological defects appearing in the form of edge dislocations \cite{mermin1979topological} of the underlying lattices, with Burgers vectors compatible with the parent PDW wave vectors, allowing localized single Cooper pairs to re-emerge. Also, such charge-$4e$ superconductivity naturally offers nontrivial pairing symmetry and exchange statistics resembling those of a topological $p+ip$ superconductor \cite{rice1995sr2ruo4, read2000paired, mackenzie2003superconductivity, luke1998time, nelson2004odd, kallin2016chiral}, yet without the necessity of broken time-reversal symmetry. These discoveries offer smoking-gun signatures and novel characterizing probes for higher-charge superconductivity and its parent PDW phases.

\begin{figure}
    \centering
    \includegraphics[width=0.45\textwidth]{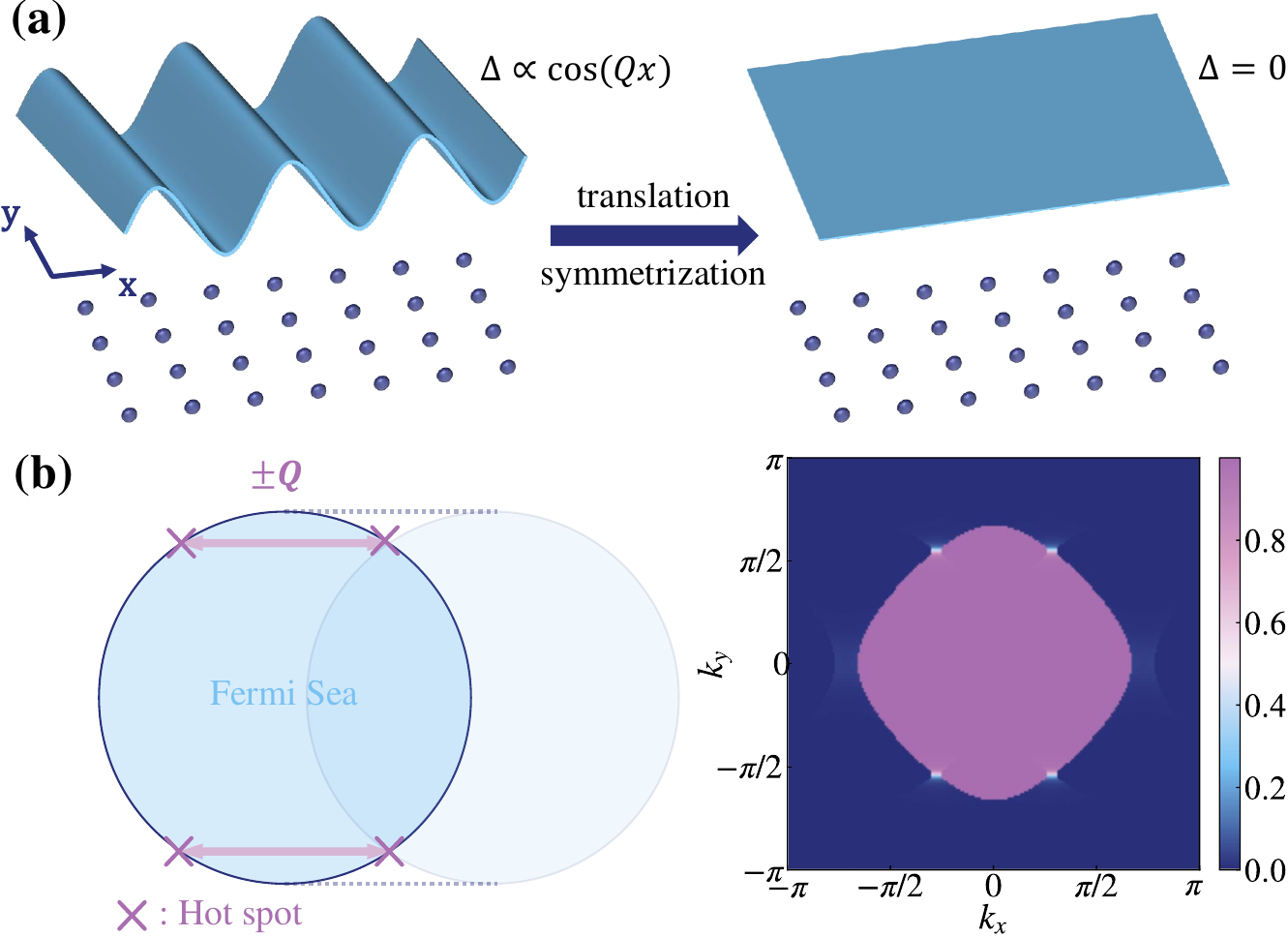}
    \caption{(a) Illustration of the model in Eq. \ref{eq:Hpdw} on a two-dimensional lattice (blue dots). In contrast to the parent state with a stripe PDW (left), the many-electron state following translation symmetrization exhibits vanishing $2e$ Cooper-pair density $\Delta$. (b) A schematic diagram illustration of the Fermi sea and the $\mathbf{\pm Q}$ momentum transfers induced due to the PDW. The hot spots (purple crosses) are located at the intersections of the original and shifted Fermi surfaces. Correspondingly, the momentum-space electron density $\langle c^\dagger_{\mathbf{k},s} c_{\mathbf{k},s}\rangle$ of the model in Eq. \ref{eq:Hpdw} clearly reveals its Fermi sea and the hot spots. Here, $V=0.1$, $Q=1.76$, and $\mu=-1.0$. }
    \label{Fig:model}
\end{figure}

\emph{Many-electron states for higher-charge superconductors.}--- For simplicity, we first start with a tight-binding model with a (mean-field) stripe PDW in 2D as our parent Hamiltonian:
\begin{eqnarray}
    \hat{H}_{\rm{PDW}}= &-&t \sum_{\langle \mathbf{r r'}\rangle, s}  (c^\dagger_{\mathbf{r'},s} c_{\mathbf{r},s} + c^\dagger_{\mathbf{r},s} c_{\mathbf{r'},s})
    - \mu \sum_{\mathbf{r}, s} c^\dagger_{\mathbf{r},s}c_{\mathbf{r},s} \nonumber\\
    &+& V\sum_{\mathbf{r}} \cos(Qx)(c^\dagger_{\mathbf{r},\uparrow}c^\dagger_{\mathbf{r},\downarrow} + {\rm h.c.}),
    \label{eq:Hpdw}
\end{eqnarray}
where $s=\uparrow,\downarrow$ labels the electron spin and $\mathbf{r}=(x,y)$ labels the sites on a $L\times L$ square lattice under periodic boundary conditions (PBCs), and we set $L=200$ unless noted otherwise; $t=1$ is the hopping amplitude as well as our unit of energy; $V$ describes the strength of a stripe PDW with wave vector $\mathbf{Q}=Q\hat{x}$ along the ${x}$-direction; see illustration in Fig. \ref{Fig:model}(a). Ideally, our choice of $Q$ should be incommensurate; however, such $Q$ would obstruct the PBC and later, the translation symmetrization in numerical models, where we are compelled to employ $Q=2\pi q/p$ with a large $p\sim O(L)$, whose distinction with an incommensurate $Q$ becomes negligible for a sufficiently large $L$. Through a particle-hole transformation $a_{\mathbf{r}, \downarrow} = c^\dagger_{\mathbf{r}, \downarrow}$ and diagonalization of $\hat H_{\rm PDW}$, we obtain the ground state $|\Psi\rangle=\prod_{\epsilon_n<0} \sum_{\mathbf{r}} [\psi_{n}(\mathbf{r},\uparrow) c^\dagger_{\mathbf{r},\uparrow} + \psi_{n}(\mathbf{r},\downarrow) a^\dagger_{\mathbf{r},\downarrow}]|\emptyset\rangle$, where $|\emptyset\rangle$ is the vacuum state of $c_\uparrow$ and $a_\downarrow$ (full state of $c_\downarrow$). On top of the electron Fermi sea dictated by the dispersion $\epsilon_\mathbf{k}=-2t(\cos k_x+\cos k_y)-\mu$, the PDW introduces coupling between $c^\dagger_{\mathbf{k},\uparrow}$ and $c^\dagger_{\mathbf{-k\pm Q},\downarrow}$ ($a_{\mathbf{k\mp Q},\downarrow}$), introducing a momentum transfer of $\pm \mathbf{Q}$, respectively; such pairing is most active around selected hot spots on the otherwise sharp Fermi surface, as shown in Fig. \ref{Fig:model}(b).

\begin{figure}
    \centering
    \includegraphics[width=0.45\textwidth]{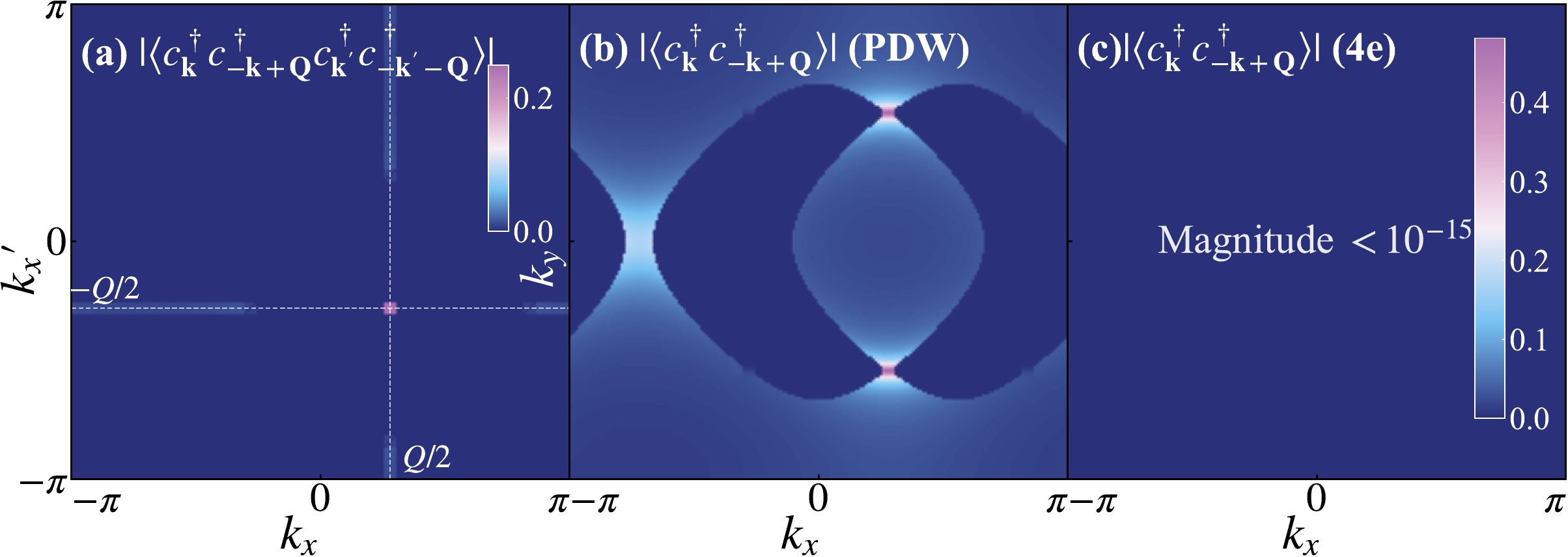}
    \caption{(a) Heat map of the four-point correlation functions $|\langle c^\dagger_{\mathbf{k}} c^\dagger_{-\mathbf{k}+\mathbf{Q}} c^\dagger_{\mathbf{k^\prime}} c^\dagger_{-\mathbf{k^\prime}-\mathbf{Q}} \rangle|$ of the charge-$4e$ superconducting state $P_T|\Psi\rangle$ show a sharp peak at $(k_x, k'_x)=(Q/2, -Q/2)$, which corresponds to the hot spots on the Fermi surface in Fig. \ref{Fig:model}(b); we have fixed $k_y=k^\prime_y=1.71$ for simpler display. (b) While the charge-$2e$ Cooper-pair density $\Delta(\mathbf{k})=\braket{c^\dagger_{\mathbf{k}} c^\dagger_{-\mathbf{k}+\mathbf{Q}}}$ flourishes in the parent PDW state $|\Psi\rangle$, (c) it is completely suppressed in the higher-charge superconducting state $P_T|\Psi\rangle$. }
    \label{Fig:4point}
\end{figure}

Next, we introduce a translation symmetrization $P_T$, where $P_T=1+T_x+T_x^2+\cdots+T_x^{L-1}$ and $T_x$ is the translation operator of one lattice spacing along the $\hat{x}$ direction. Such a $P_T$ is sufficient for restoring the translation symmetry broken by the stripe PDW in Eq. \ref{eq:Hpdw} and straightforwardly generalizable in more complex scenarios, some of which we will discuss later and in Ref. \cite{supp}. The subsequent projected wavefunction, $P_T|\Psi\rangle = T_x (P_T|\Psi\rangle)$, is translation invariant, just as the vestigial order after the (partial) melting of the initial PDW. Interestingly, unlike the parent PDW state $|\Psi\rangle$ (Fig. \ref{Fig:4point}(b) \footnote{We note that the PDW has a finite charge-$2e$ Cooper-pair density in most places, yet the average of states is different from the average of expectation values. }), the Cooper-pair expectation values, $\langle c^\dagger_{\mathbf{r}, \uparrow} c^\dagger_{\mathbf{r'}, \downarrow}\rangle$, vanish universally for arbitrary $\mathbf{r}$ and $\mathbf{r'}$, showing that the single $2e$ Cooper-pair sector is fully absent in $P_T|\Psi\rangle$; see Fig. \ref{Fig:4point}(c). Further, we can fully suppress the charge-$2e$ defining feature, the long-range pair-pair correlations, by introducing a finite coherence length to the parent PDWs; see Ref. \cite{supp} for details. Instead, the minimum charge difference is $4e$ - two Cooper pairs, as shown by the finite expectation value of $\langle c^\dagger_{\mathbf{k},\uparrow} c^\dagger_{\mathbf{-k+Q},\downarrow} c^\dagger_{\mathbf{k'},\uparrow} c^\dagger_{\mathbf{-k'-Q},\downarrow} \rangle$ in Fig. \ref{Fig:4point}(a). We detail our method for evaluating such expectation values in the Supplemental Material \cite{supp}. Such a contrast between two-electrons and four-electrons terms already indicates that $P_T|\Psi\rangle$ deviates significantly from a simple Hartree-Fock state, whose correlations always follow straightforward factorizations of Wick's theorem.

The Gutzwiller projection of spinful electrons offers quantum spin-liquid wavefunctions for variational studies and universal characterizations \cite{yokoyama1987variational1, yokoyama1987variational2, gros1989physics, wen1991mean, paramekanti2001projected, himeda2002stripe, motrunich2005variational, edegger2007gutzwiller,zhang2011entanglement, zhang2012quasiparticle,grover2013entanglement,zhang2013establishing, iqbal2011projected, iqbal2013gapless}, by retaining the physically relevant spinon degrees of freedom while removing the chargon degrees of freedom. Effectively, the projection binds the spin-up electrons and spin-down holes as an auxiliary (gauge) field. Likewise, a stripe PDW consists of two order parameter fields $\Delta_\mathbf{Q}(\mathbf{r}), \Delta_{\mathbf{-Q}}(\mathbf{r}) \in \mathbb{C}$, corresponding to its $\mathbf{\pm Q}$ components, whose anti-binding $\Delta^*_\mathbf{Q}(\mathbf{r})\Delta_{\mathbf{-Q}}(\mathbf{r})$ and binding $\Delta_\mathbf{Q}(\mathbf{r})\Delta_{\mathbf{-Q}}(\mathbf{r})$ are associated with the CDW and the charge-$4e$ superconductivity, breaking translation symmetry and charge conservation, respectively - the translation symmetrization projects out the former and retains the latter degrees of freedom. Phenomenologically, the projection $P_T$ effectively integrates over the relative spatial phases of the PDW order parameters. This process fundamentally corresponds to the proliferation of phase fluctuations—-dual to vortices—-that melts the single-Cooper-pair order while binding the two order parameter fields, consistent with previous Ginzburg-Landau theories \cite{agterberg2008dislocations, berg2009charge, herland2010phase, agterberg2011conventional, jian2021charge, fernandes2021charge, lin2025theory}. Although a local 'vortex density' operator is not directly definable within this purely electronic many-body wavefunction, this vortex-proliferation picture can be indirectly visualized through topological lattice defects, as we will demonstrate later.

\begin{figure}
    \centering
    \includegraphics[width=0.45\textwidth]{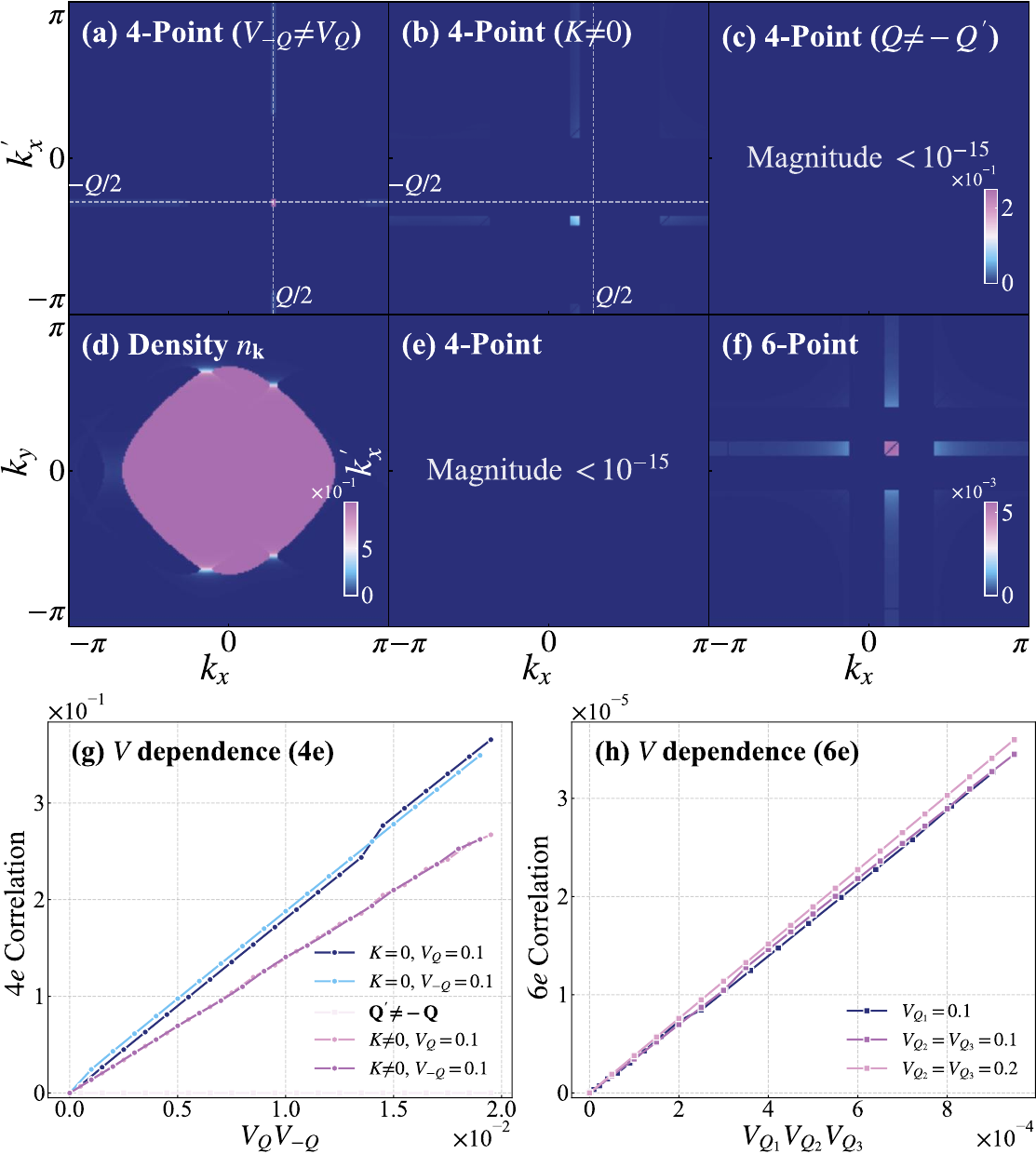}
    \caption{The correlations $|\langle c^\dagger_{\mathbf{k}} c^\dagger_{-\mathbf{k}+\mathbf{Q}} c^\dagger_{\mathbf{k^\prime}} c^\dagger_{-\mathbf{k^\prime}-\mathbf{Q}} \rangle|$ of the $P_T|\Psi \rangle$ states with (a) asymmetric pairing amplitude $V_\mathbf{Q}\ne V_{\mathbf{-Q}}$, (b) asymmetric Fermi sea with non-zero total momentum $\mathbf{K}$, and (c) nonreciprocal PDW momentum $\mathbf{Q'}\ne-\mathbf{Q}$, suggests that the charge-$4e$ superconductivity may exists as long as momentum conserves despite broken inversion symmetry. (d) The electron density $\langle c^\dagger_{\mathbf{k},s} c_{\mathbf{k},s}\rangle$ of a charge-$6e$ superconductor displays hot spots on its Fermi surface. The corresponding $P_T|\Psi\rangle$ exhibits vanishing charge-$2e$ and (e) charge-$4e$ expectation values, yet (f) a sharp signal in charge-$6e$ correlations (we have fixed $k_y=k^\prime_y=k^{\prime\prime}_y=2.0$). (g) The charge-$4e$ order parameter $V_{4e}$ scales proportionally to both $V_\mathbf{Q}$ and $V_{\mathbf{-Q}}$; (h) similarly,  the charge-$6e$ order parameter scales tri-linearly to each of the three parent PDWs' amplitudes. Here, we set $L=1000$. }
    \label{Fig:Generalization}
\end{figure}

Intuitively, we may understand the above unusual properties of $P_T |\Psi\rangle$ through the perturbation theory for relatively small $V$. The zeroth-order state $|\Psi_0 \rangle$, the electron Fermi sea of the first line in Eq. \ref{eq:Hpdw}, is fully translation-symmetric with a zero total momentum, thus remains intact upon $P_T$. Following first-order perturbation, the contributions to the state, $|\Psi_1 \rangle \propto V \sum_\mathbf{k} (c^\dagger_{\mathbf{k}, \uparrow}c^\dagger_{\mathbf{-k\pm Q}, \downarrow} + \mbox{h.c.})|\Psi_0 \rangle$, introduce charge-$2e$ Cooper pairs, yet alter the total momentum by $\mathbf{\pm Q}$, and subsequently, become projected out by $P_T$. It is not until the second order that  a nontrivial contribution $|\Psi_2 \rangle \propto V^2 \sum_{\mathbf{k},\mathbf{k'}} c^\dagger_{\mathbf{k}, \uparrow}c^\dagger_{\mathbf{-k\pm Q}, \downarrow} \cdot c^\dagger_{\mathbf{k'}, \uparrow}c^\dagger_{\mathbf{-k'\mp Q}, \downarrow} |\Psi_0 \rangle$ emerges. This contribution corresponds to sequential scatterings from the PDW with with opposing, and thus mutually canceling, momentum transfers, enabling the state to conserve its total momentum and survive the projection $P_T$. Indeed, our numerical results indicate that $P_T|\Psi\rangle$'s expectation values of charge-$4e$ Cooper pairs are most significant at momenta $\mathbf{\pm Q}$: $\langle c^\dagger_{\mathbf{k}, \uparrow}c^\dagger_{\mathbf{-k+Q}, \downarrow}c^\dagger_{\mathbf{k'}, \uparrow}c^\dagger_{\mathbf{-k'-Q}, \downarrow}\rangle$, with amplitudes quadratic in $V$, see Fig. \ref{Fig:Generalization}(g).

Importantly, while previous Ginzburg-Landau theories often incorporate point-group symmetries -- inversion symmetry for charge-$4e$ superconductivity \cite{agterberg2008dislocations, berg2009charge, herland2010phase, jian2021charge, fernandes2021charge}, $C_3$ rotation symmetry for charge-$6e$ superconductivity \cite{agterberg2011conventional, lin2025theory} -- to stabilize and ensure the degeneracy of coexisting PDW components, our quantum many-electron states explicitly demonstrate that momentum conservation alone is sufficient to drive the subsequent emergence of higher-charge condensates. For instance, we may generalize $P_T|\Psi\rangle$ to cases without inversion symmetry, e.g., either a non-reciprocal PDW $V_\mathbf{Q} \sum_\mathbf{k}c^\dagger_{\mathbf{k}, \uparrow}c^\dagger _{\mathbf{-k+Q}, \downarrow} + V_{\mathbf{Q'}} \sum_\mathbf{k} c^\dagger_{\mathbf{k}, \uparrow}c^\dagger_{\mathbf{-k+Q'}, \downarrow} + \mbox{h.c.}$ ($\mathbf{Q'}=-\mathbf{Q}$), or more fundamentally, an asymmetric Fermi sea $|\Psi_0\rangle$. Note that in the latter case, the translation symmetrization should be adjusted as $P_T=1+T_x e^{iK}+T_x^2 e^{i2K}+\cdots+T_x^{L-1} e^{iK(L-1)}$ for a Fermi sea with a finite total momentum $\mathbf{K}$; for simplicity, here we consider $\mathbf{K}=K\hat{x}$. Consistent with the aforementioned suppression of charge-$2e$ sectors, Figs. \ref{Fig:Generalization}(a)(b) demonstrate that a finite charge-$4e$ signal is clearly evident in both scenarios. Further, the charge-$4e$ amplitude at $\langle c^\dagger_{\mathbf{k}, \uparrow}c^\dagger_{\mathbf{-k+Q}, \downarrow}c^\dagger_{\mathbf{k'}, \uparrow}c^\dagger_{\mathbf{-k'-Q}, \downarrow}\rangle$, which we dub the charge-$4e$ order parameter $V_{4e}$ and summarize in Fig. \ref{Fig:Generalization}(g), shows a clear linear dependence on both $V_\mathbf{Q}$ and $V_{\mathbf{-Q}}$: $V_{4e}\propto V_\mathbf{Q}V_{\mathbf{-Q}}$, consistent with our perturbative analysis and irrespective of broken inversion symmetry. On the contrary, all charge-$4e$ expectation values $\langle c^\dagger_{\mathbf{r},\uparrow} c^\dagger_{\mathbf{r'},\downarrow} c^\dagger_{\mathbf{r''},\uparrow} c^\dagger_{\mathbf{r'''},\downarrow}\rangle$ vanish in $P_T|\Psi\rangle$ in the absence of momentum conservation, e.g., the momentum components $\mathbf{Q'}\ne-\mathbf{Q}$ within the PDW, as we summarize in Figs. \ref{Fig:Generalization}(c)(g). These results and conclusions also extend to relatively large values of $V_{\mathbf{\pm Q}}$, as we discuss further in the Supplemental Material \cite{supp}.

Further, we can generalize such a construction to higher-charge superconductivity, e.g., charge-$6e$ Cooper pairs with three PDW momentum components, $\mathbf{Q_1}$, $\mathbf{Q_2}$, and $\mathbf{Q_3}$, which conserves total momentum $\mathbf{Q_1+Q_2+Q_3}=0$. For instance, we employ the following PDW term on top of the Fermi sea (first line) in Eq. \ref{eq:Hpdw}:
\begin{eqnarray}
H_{{\rm PDW}} &=&  \sum_\mathbf{r} V(x) c^\dagger_{\mathbf{r},\uparrow} c^\dagger_{\mathbf{r},\downarrow} + {\rm h.c.}, \nonumber \\
V(x) &=& V_\mathbf{Q} e^{iQx}+ V_{-\mathbf{Q}/2}e^{-iQx/2}.
\label{eq:V6e}
\end{eqnarray}
Following translation symmetrization, the resulting state $P_T|\Psi\rangle$ witnesses vanishing charge-$2e$ and charge-$4e$ correlations, yet a clear charge-$6e$ pairing signature ($\langle c^\dagger_{\mathbf{k}, \uparrow}c^\dagger_{\mathbf{-k+Q}, \downarrow}c^\dagger_{\mathbf{k'}, \uparrow}c^\dagger_{\mathbf{-k'-Q}/2, \downarrow}c^\dagger_{\mathbf{k''}, \uparrow}c^\dagger_{\mathbf{-k''-Q}/2, \downarrow}\rangle$), which scales as $V^3$, or proportionally to the amplitude of each of the participating PDWs. These results are presented in Figs. \ref{Fig:Generalization}(d)-(f)(h). We note that potential signatures on sequential or coexisting charge-$4e$ and charge-$6e$ orderings observed in CsV$_3$Sb$_5$ materials \cite{ge2024charge} are consistent with the previously hosting PDW patterns \cite{chen2021roton, zhao2021cascade}, which conserve the total momentum under both charge-$4e$ and charge-$6e$ mechanisms. Generalizations to higher dimensions or other PDW scenarios, such as the checkerboard order, are straightforward, albeit a corresponding translation symmetrization $P_T$ may need to be introduced.

\begin{figure}
    \centering
    \includegraphics[width=0.5\textwidth]{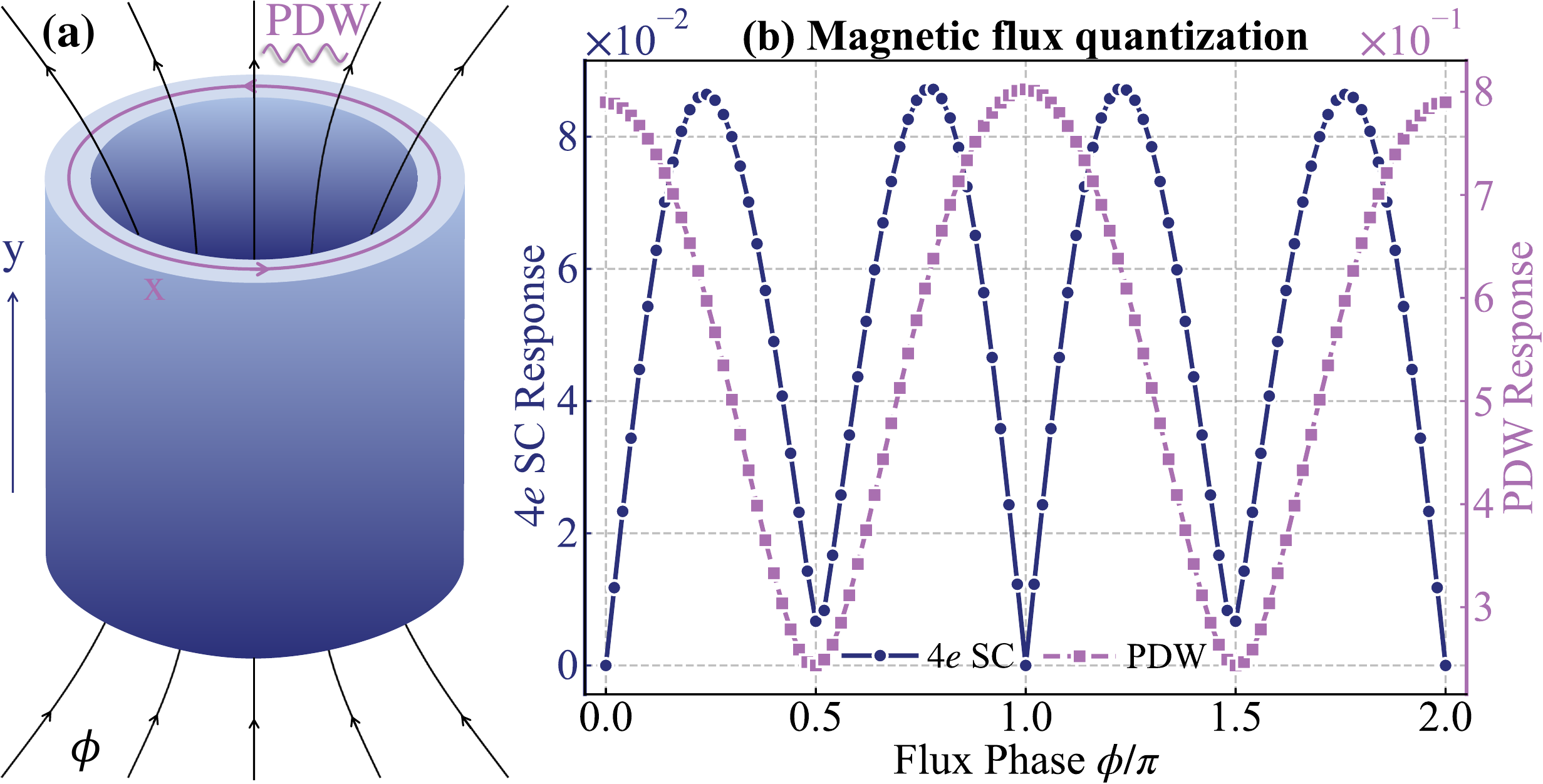}
    \caption{(a) An illustration of our interferometer setup $P_T|\Psi(\phi)\rangle$: translation-symmetrizing the parent PDW model in Eq. \ref{eq:Hpdw} with a PDW in the $x$-direction and a magnetic flux $\phi$ in the $y$-direction. (b) Unlike the parent PDW system $|\Psi(\phi)\rangle$ with a $\pi$ periodicity in $\phi$ corresponding to conventional charge-$2e$ Cooper pairs, the susceptibility $\mathcal{L(\phi)}$ in Eq. \ref{eq:susceptibility} of $P_T|\Psi(\phi)\rangle$ exhibits an unmistakable charge-$4e$ signature: $\pi/2$ period with respect to the threaded magnetic flux $\phi$. Here, our system size is $L=1000$. }
    \label{Fig:LittlePark}
\end{figure}

\emph{Phenomena of higher-charge superconductivity.}--- Our microscopic many-electron models of higher-charge superconductors have opened new opportunities to understand and discover their novel properties. For instance, we may probe their exotic magnetic-flux quantization by measuring the periodicity of their physical properties in an interferometer geometry. In particular, we may introduce to the parent PDW model in Eq. \ref{eq:Hpdw} a magnetic flux along the $y$-direction through a phase $\phi$ across the boundary along the $x$-direction; see our setup geometry in Fig. \ref{Fig:LittlePark}(a). We neglect the Zeeman effect on the electron spins and the magnetic flux vortices for simplicity.

After the translation symmetrization, fully compatible with the magnetic flux, we evaluate the susceptibility of the charge-$4e$ Cooper pair:
\begin{eqnarray}
\mathcal{L(\phi)}=\frac{\left.\partial \langle c^\dagger_{\mathbf{k}, \uparrow}c^\dagger_{\mathbf{-k+Q}, \downarrow}c^\dagger_{\mathbf{k'}, \uparrow}c^\dagger_{\mathbf{-k'-Q}, \downarrow}\rangle\right|_{P_T|\Psi(\phi)\rangle} }{\partial V},
\label{eq:susceptibility}
\end{eqnarray}
for the subsequent many-electron state $P_T|\Psi(\phi)\rangle$, which we summarize in Fig. \ref{Fig:LittlePark}(b). Indeed, $\mathcal{L(\phi)}$ exhibits clear, distinctive $\pi/2$ periodicity in magnetic fluxes and offers strong evidence for its charge-$4e$ fundamentals. We attribute the slight deviations from a perfect $4e$-period to residue contributions of the electrons, especially the remaining Fermi surface, and the finite-size effects; please refer to the Supplemental Material for additional results and generalizations \cite{supp}. This is in sharp contrast with its parent PDW state, where the $\pi$ magnetic-flux period plays a dominant role in the pairing susceptibility contributed by charge-$2e$ Cooper pairs.

In addition, our formalism may reveal new features of higher-charge superconducting states: while the charge-$2e$ Cooper-pair channel is initially gapped out, localized Cooper pairs may re-emerge at certain topological defects, as long as their Burgers vector $\mathbf{B}$ is compatible, i.e., parallel or anti-parallel, with the parent PDW wave vector $\mathbf{Q}$. Intuitively, a topological defect may serve as a source or drain of momentum along its Burgers vector \cite{supp}, thus assisting charge-$2e$ Cooper pairs originating from single PDW scatterings to conserve momentum and exist locally. We note that a topological defect and its Burgers vector - essentially a topological invariant - remains robust upon perturbations; likewise, such localized Cooper pairs also remain relatively stable. When and only when two topological defects with opposing Burgers vectors encounter and annihilate, their respective local charge-$2e$ Cooper pairs with $\mathbf{\pm Q}$ momentum merge into the charge-$4e$ Cooper pair condensate.

\begin{figure}
    \centering
    \includegraphics[width=0.45\textwidth]{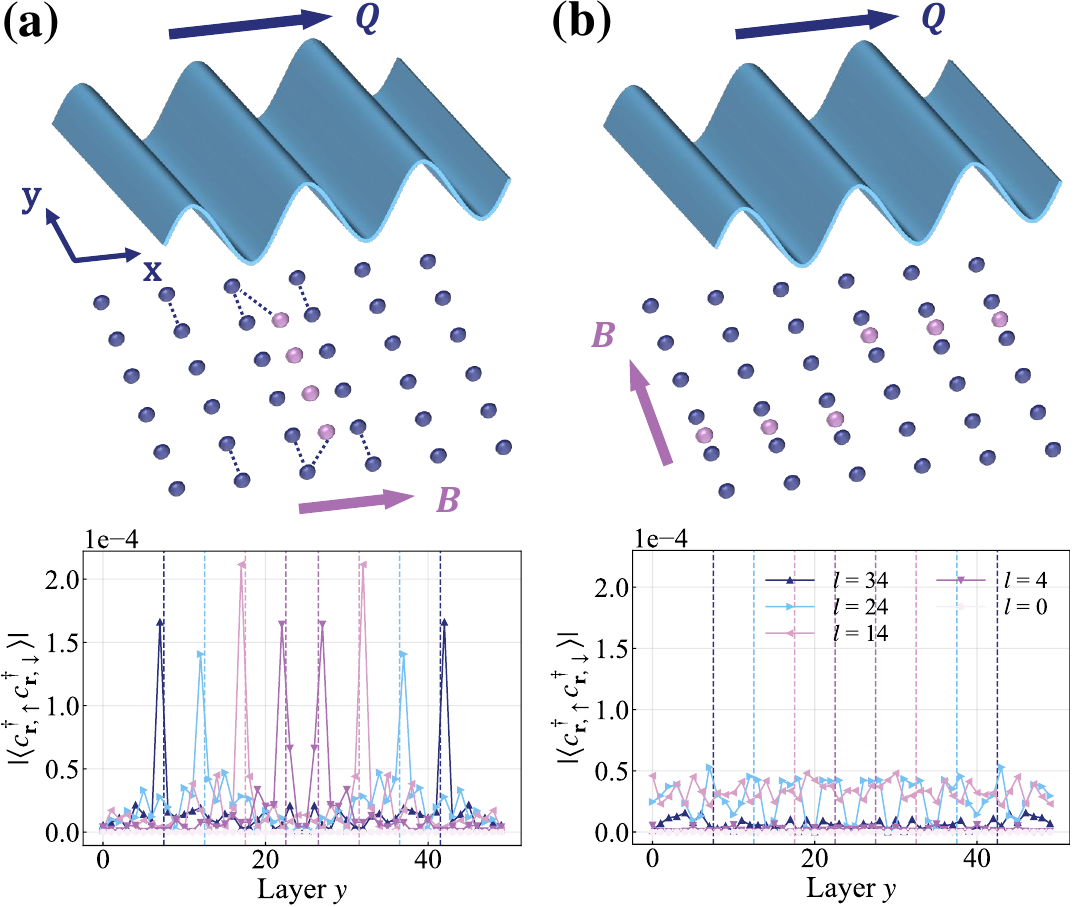}
    \caption{(a) Charge-$2e$ Cooper pairs re-emerge locally at topological defects of the underlying lattice when the Burgers vector $\mathbf{B}$ aligns with the parent PDW wave vector $\pm \mathbf{Q}$, yet (b) no such modes are visible when $\mathbf{B}$ and $\mathbf{Q}$ are vertical. The upper panels show schematic illustrations of the lattices with a pair of topological defects and the PDWs before translation symmetrization $P_T$. The dashed lines illustrate the local hopping connectivity, which becomes 1-to-2 on the edge-dislocation sites. The lower panels are the charge-$2e$ expectation values $\langle c^\dagger_{\mathbf{r},\uparrow} c^\dagger_{\mathbf{r},\downarrow} \rangle$ for such $P_T|\Psi \rangle$ with various distances $l$ between the two topological defects. Here, we have $L=50$. }
    \label{Fig:dislocation}
\end{figure}

To demonstrate such localized charge-$2e$ Cooper pairs, we introduce a pair of edge dislocations with opposing Burgers vectors $\pm\mathbf{B}$ into the 2D lattice hosting the parent Hamiltonian in Eq. \ref{eq:Hpdw}, see Fig. \ref{Fig:dislocation} upper panels. In particular, we insert $l$ additional lattice sites in the bulk, thus separating the two edge dislocations by a distance $l$. Except in the vicinity of the topological defects, the system remains effectively translation-invariant, and we can generalize our translation symmetrization $P_T$ accordingly, see further details in Ref. \cite{supp}.

As a result, Fig. \ref{Fig:dislocation} lower panels display the charge-$2e$ Cooper pair density $\langle c^\dagger_{\mathbf{r},\uparrow} c^\dagger_{\mathbf{r},\downarrow} \rangle$ for the many-electron state $P_T|\Psi\rangle$. When the Burgers vector $\bf{B}$ is perpendicular to the PDW wave vector $\bf{Q}$, charge-$2e$ Cooper pairs remain fully absent. In contrast, when $\bf{B}$ is parallel to $\bf{Q}$, clear charge-$2e$ pairing signals emerge at the dislocations, while everywhere else in the lattice remains oblivious to the topological defects and minimally charge-$4e$. From the perspective of phase fluctuations, such a topological defect breaks translation symmetry and acts as a local ``vacuum" for the proliferating vortices, allowing the charge-2e Cooper pairs to re-emerge nearby. As we gradually decrease the separation $l$ between the two dislocations, the localized charge-$2e$ Cooper pairs approach and eventually annihilate each other. During this process, the pinned vortices re-saturate the former defect cores, causing the local charge-2e signals to vanish and fully restoring the translation-symmetric, vestigial charge-4e superconducting state without topological defects. Similar conclusions also hold for open boundaries of higher-charge superconductors, which, when normal to $\bf{Q}$, may transfer momentum and thus host localized charge-$2e$ Cooper pairs, as we further discuss in Ref. \cite{supp}. These unusual localized-Cooper-pair phenomena are fully consistent with our theoretical expectations, and offer smoking-gun signatures for identifying charge-$4e$ superconductivity as well as parent PDW in candidate materials, e.g., CsV$_3$Sb$_5$, via scanning tunneling spectroscopy experiments.


\emph{Discussion.}--- We have established many-electron characterizations of higher-charge superconductivity by translation-symmetrizing parent PDW systems, consistent with the Ginzburg-Landau theory of vestigial orders. In particular, with simple stripe PDWs, we clearly demonstrate charge-$4e$ and charge-$6e$ superconductivity, with vanishing charge-$2e$ Cooper-pair sectors as well as long-range pair-pair correlations \cite{supp}. Interestingly, momentum conservation, rather than stricter point-group symmetries, is in principle sufficient for their establishment and thus significantly broadens their potential material platforms.

Beyond serving as a conceptual framework, our many-electron formalism provides concrete predictions for experimental observables. The obtained higher-charge superconductors exhibit consistent magnetic flux quantization and periodicity in their interferometry. We also show that charge-$2e$ Cooper pairs may re-emerge at topological defects of the underlying lattice, if their Burgers vector and the parent PDW wave vector are compatible. In addition, as a potential manifestation under specific settings (e.g., complex hopping amplitudes), the spin-up (or spin-down) electrons of the charge-$4e$ superconductors can exhibit topological exchange statistics that resemble a nodeless $p+ip$-wave superconductor, even while the overall system preserves macroscopic time-reversal symmetry. This feature thus offers a novel arena for nontrivial topological phenomena and applications, as we discuss further in the Supplemental Material \cite{supp}. Further theoretical insights from the perspective of topological quantum field theory are detailed in Refs. \cite{gao2025topological, shi2025non}.Overall, our formalism offers unprecedented opportunities to understand and identify these exotic quantum matters.

Finally, generalizations of the current translation-symmetrization formalism to higher dimensions, other density-wave scenarios, and alternative parent numerical ansatzes, such as matrix product states and quantum Monte Carlo methods, are straightforward. Interestingly, applications to spin density waves and charge density waves may offer a novel perspective and method for the physics of intertwined orders \cite{davis2013concepts, wang2014charge, keimer2015quantum, fernandes2019intertwined} in strongly correlated systems \cite{gruner1988dynamics, gruner1994dynamics}. There, in addition to the universal behaviors of the superconducting phases that we have mainly focused on, details such as the parent density wave's pattern, strength, and wave vector may impact their overall energetic competitiveness. Besides, as in the topological theories of gapped quantum spin liquids, it will be crucial to investigate the physics of such Cooper-pair-binding vortices, based on particular Fermi seas entirely gapped out by the parent density waves and thus clean from residual contributions.

\emph{Acknowledgment:} We acknowledge Steven A. Kivelson, Fa Wang, Jian Wang, Xi Dai, and Hong-Hao Tu for helpful discussions and support from the National Key R\&D Program of China (Grant No.2021YFA1401900 \& No.2022YFA1403700), the National Natural Science Foundation of China (Grants No.12174008 \& No.92270102), and Shanghai Municipal Science and Technology Project (Grant No.25LZ2601100).

\bibliography{ref.bib}

@article{van2010discovery,
  title={The discovery of superconductivity},
  author={Van Delft, Dirk and Kes, Peter},
  journal={Physics today},
  volume={63},
  number={9},
  pages={38--43},
  year={2010},
  publisher={AIP Publishing}
}

@article{bardeen1957theory,
  title={Theory of superconductivity},
  author={Bardeen, John and Cooper, Leon N and Schrieffer, John Robert},
  journal={Physical review},
  volume={108},
  number={5},
  pages={1175},
  year={1957},
  publisher={APS}
}

@article{chen2021roton,
  title={Roton pair density wave in a strong-coupling kagome superconductor},
  author={Chen, Hui and Yang, Haitao and Hu, Bin and Zhao, Zhen and Yuan, Jie and Xing, Yuqing and Qian, Guojian and Huang, Zihao and Li, Geng and Ye, Yuhan and others},
  journal={Nature},
  volume={599},
  number={7884},
  pages={222--228},
  year={2021},
  publisher={Nature Publishing Group UK London}
}

@article{agterberg2020physics,
  title={The physics of pair-density waves: cuprate superconductors and beyond},
  author={Agterberg, Daniel F and Davis, JC S{\'e}amus and Edkins, Stephen D and Fradkin, Eduardo and Van Harlingen, Dale J and Kivelson, Steven A and Lee, Patrick A and Radzihovsky, Leo and Tranquada, John M and Wang, Yuxuan},
  journal={Annual Review of Condensed Matter Physics},
  volume={11},
  number={1},
  pages={231--270},
  year={2020},
  publisher={Annual Reviews}
}

@article{fradkin2015colloquium,
  title={Colloquium: Theory of intertwined orders in high temperature superconductors},
  author={Fradkin, Eduardo and Kivelson, Steven A and Tranquada, John M},
  journal={Reviews of Modern Physics},
  volume={87},
  number={2},
  pages={457--482},
  year={2015},
  publisher={APS}
}

@article{berg2009charge,
  title={Charge-{$4e$} superconductivity from pair-density-wave order in certain high-temperature superconductors},
  author={Berg, Erez and Fradkin, Eduardo and Kivelson, Steven A},
  journal={Nature Physics},
  volume={5},
  number={11},
  pages={830--833},
  year={2009},
  publisher={Nature Publishing Group UK London}
}

@article{jian2021charge,
  title={Charge-{$4e$} superconductivity from nematic superconductors in two and three dimensions},
  author={Jian, Shao-Kai and Huang, Yingyi and Yao, Hong},
  journal={Physical review letters},
  volume={127},
  number={22},
  pages={227001},
  year={2021},
  publisher={APS}
}

@article{agterberg2008dislocations,
  title={Dislocations and vortices in pair-density-wave superconductors},
  author={Agterberg, DF and Tsunetsugu, H},
  journal={Nature Physics},
  volume={4},
  number={8},
  pages={639--642},
  year={2008},
  publisher={Nature Publishing Group UK London}
}

@article{agterberg2011conventional,
  title={Conventional and charge-six superfluids from melting hexagonal Fulde-Ferrell-Larkin-Ovchinnikov phases in two dimensions},
  author={Agterberg, DF and Geracie, M and Tsunetsugu, H},
  journal={Physical Review B—Condensed Matter and Materials Physics},
  volume={84},
  number={1},
  pages={014513},
  year={2011},
  publisher={APS}
}

@article{nie2014quenched,
  title={Quenched disorder and vestigial nematicity in the pseudogap regime of the cuprates},
  author={Nie, Laimei and Tarjus, Gilles and Kivelson, Steven Allan},
  journal={Proceedings of the National Academy of Sciences},
  volume={111},
  number={22},
  pages={7980--7985},
  year={2014},
  publisher={National Academy of Sciences}
}

@article{ge2024charge,
  title={Charge-{$4e$} and charge-{$6e$} flux quantization and higher charge superconductivity in kagome superconductor ring devices},
  author={Ge, Jun and Wang, Pinyuan and Xing, Ying and Yin, Qiangwei and Wang, Anqi and Shen, Jie and Lei, Hechang and Wang, Ziqiang and Wang, Jian},
  journal={Physical Review X},
  volume={14},
  number={2},
  pages={021025},
  year={2024},
  publisher={APS}
}

@article{zhou2022chern,
  title={Chern Fermi pocket, topological pair density wave, and charge-{$4e$} and charge-{$6e$} superconductivity in kagom{\'e} superconductors},
  author={Zhou, Sen and Wang, Ziqiang},
  journal={Nature Communications},
  volume={13},
  number={1},
  pages={7288},
  year={2022},
  publisher={Nature Publishing Group UK London}
}

@article{liu2023charge,
  title={Charge-{$4e$} superconductivity and chiral metal in 45-twisted bilayer cuprates and related bilayers},
  author={Liu, Yu-Bo and Zhou, Jing and Wu, Congjun and Yang, Fan},
  journal={Nature Communications},
  volume={14},
  number={1},
  pages={7926},
  year={2023},
  publisher={Nature Publishing Group UK London}
}

@article{gnezdilov2022solvable,
  title={Solvable model for a charge-{$4e$} superconductor},
  author={Gnezdilov, Nikolay V and Wang, Yuxuan},
  journal={Physical Review B},
  volume={106},
  number={9},
  pages={094508},
  year={2022},
  publisher={APS}
}

@article{jiang2017charge,
  title={Charge-{$4e$} superconductors: A Majorana quantum Monte Carlo study},
  author={Jiang, Yi-Fan and Li, Zi-Xiang and Kivelson, Steven A and Yao, Hong},
  journal={Physical Review B},
  volume={95},
  number={24},
  pages={241103},
  year={2017},
  publisher={APS}
}

@article{soldini2024charge,
  title={Charge-{$4e$} superconductivity in a Hubbard model},
  author={Soldini, Martina O and Fischer, Mark H and Neupert, Titus},
  journal={Physical Review B},
  volume={109},
  number={21},
  pages={214509},
  year={2024},
  publisher={APS}
}

@article{fernandes2021charge,
  title={Charge-{$4e$} superconductivity from multicomponent nematic pairing: Application to twisted bilayer graphene},
  author={Fernandes, Rafael M and Fu, Liang},
  journal={Physical review letters},
  volume={127},
  number={4},
  pages={047001},
  year={2021},
  publisher={APS}
}

@article{wu2024d,
  title={d-wave charge-{$4e$} superconductivity from fluctuating pair density waves},
  author={Wu, Yi-Ming and Wang, Yuxuan},
  journal={npj Quantum Materials},
  volume={9},
  number={1},
  pages={66},
  year={2024},
  publisher={Nature Publishing Group UK London}
}

@article{lin2025theory,
  title={Theory of the charge-{$6e$} condensed phase in kagome-lattice superconductors},
  author={Lin, Tong-Yu and Song, Feng-Feng and Zhang, Guang-Ming},
  journal={Physical Review B},
  volume={111},
  number={5},
  pages={054508},
  year={2025},
  publisher={APS}
}

@article{zeng2024high,
  title={High-order time-reversal symmetry breaking normal state},
  author={Zeng, Meng and Hu, Lun-Hui and Hu, Hong-Ye and You, Yi-Zhuang and Wu, Congjun},
  journal={Science China Physics, Mechanics \& Astronomy},
  volume={67},
  number={3},
  pages={237411},
  year={2024},
  publisher={Springer}
}

@article{pan2024frustrated,
  title={Frustrated superconductivity and sextetting order},
  author={Pan, Zhiming and Lu, Chen and Yang, Fan and Wu, Congjun},
  journal={Science China Physics, Mechanics \& Astronomy},
  volume={67},
  number={8},
  pages={287412},
  year={2024},
  publisher={Springer}
}

@article{liu2024nematic,
  title={Nematic superconductivity and its critical vestigial phases in the quasicrystal},
  author={Liu, Yu-Bo and Zhou, Jing and Yang, Fan},
  journal={Physical Review Letters},
  volume={133},
  number={13},
  pages={136002},
  year={2024},
  publisher={APS}
}

@article{laughlin1983anomalous,
  title={Anomalous quantum Hall effect: An incompressible quantum fluid with fractionally charged excitations},
  author={Laughlin, Robert B},
  journal={Physical Review Letters},
  volume={50},
  number={18},
  pages={1395},
  year={1983},
  publisher={APS}
}

@article{wen1991mean,
  title={Mean-field theory of spin-liquid states with finite energy gap and topological orders},
  author={Wen, Xiao-Gang},
  journal={Physical Review B},
  volume={44},
  number={6},
  pages={2664},
  year={1991},
  publisher={APS}
}

@article{herland2010phase,
  title={Phase transitions in a three dimensional {U(1)$\times$U (1)} lattice London superconductor: Metallic superfluid and charge-{$4e$} superconducting states},
  author={Herland, Egil V and Babaev, Egor and Sudb{\o}, Asle},
  journal={Physical Review B—Condensed Matter and Materials Physics},
  volume={82},
  number={13},
  pages={134511},
  year={2010},
  publisher={APS}
}

@article{meissner1933neuer,
  title={Ein neuer effekt bei eintritt der supraleitf{\"a}higkeit},
  author={Meissner, Walther and Ochsenfeld, Robert},
  journal={Naturwissenschaften},
  volume={21},
  number={44},
  pages={787--788},
  year={1933},
  publisher={Springer-Verlag Berlin/Heidelberg}
}

@article{fulde1964superconductivity,
  title={Superconductivity in a strong spin-exchange field},
  author={Fulde, Peter and Ferrell, Richard A},
  journal={Physical Review},
  volume={135},
  number={3A},
  pages={A550},
  year={1964},
  publisher={APS}
}

@article{larkin1965nonuniform,
  title={Nonuniform state of superconductors},
  author={Larkin, AI and Ovchinnikov, Yu N},
  journal={Soviet Physics-JETP},
  volume={20},
  number={3},
  pages={762--762},
  year={1965}
}

@article{cooper1956bound,
  title={Bound electron pairs in a degenerate Fermi gas},
  author={Cooper, Leon N},
  journal={Physical Review},
  volume={104},
  number={4},
  pages={1189},
  year={1956},
  publisher={APS}
}

@incollection{ginzburg2009theory,
  title={On the theory of superconductivity},
  author={Ginzburg, Vitaly L and Landau, Lev D},
  booktitle={On superconductivity and superfluidity: a scientific autobiography},
  pages={113--137},
  year={2009},
  publisher={Springer}
}

@article{onnes1911resistance,
  title={The resistance of pure mercury at helium temperatures},
  author={Onnes, Kamerlingh},
  journal={Commun. Phys. Lab. Univ. Leiden, b},
  volume={120},
  year={1911}
}

@article{wu2023pair,
  title={Pair density wave order from electron repulsion},
  author={Wu, Yi-Ming and Nosov, Pavel A and Patel, Aavishkar A and Raghu, S},
  journal={Physical Review Letters},
  volume={130},
  number={2},
  pages={026001},
  year={2023},
  publisher={APS}
}

@article{ticea2024pair,
  title={Pair density wave order in multiband systems},
  author={Ticea, Nicole S and Raghu, Srinivas and Wu, Yi-Ming},
  journal={Physical Review B},
  volume={110},
  number={9},
  pages={094515},
  year={2024},
  publisher={APS}
}

@article{hamidian2016detection,
  title={Detection of a Cooper-pair density wave in {Bi$_2$Sr$_2$CaCu$_2$O$_8+x$}},
  author={Hamidian, MH and Edkins, Stephen David and Joo, Sang Hyun and Kostin, A and Eisaki, H and Uchida, S and Lawler, MJ and Kim, E-A and Mackenzie, Andrew P and Fujita, K and others},
  journal={Nature},
  volume={532},
  number={7599},
  pages={343--347},
  year={2016},
  publisher={Nature Publishing Group UK London}
}

@article{liu2021discovery,
  title={Discovery of a Cooper-pair density wave state in a transition-metal dichalcogenide},
  author={Liu, Xiaolong and Chong, Yi Xue and Sharma, Rahul and Davis, JC S{\'e}amus},
  journal={Science},
  volume={372},
  number={6549},
  pages={1447--1452},
  year={2021},
  publisher={American Association for the Advancement of Science}
}

@article{gu2023detection,
  title={Detection of a pair density wave state in {UTe$_2$}},
  author={Gu, Qiangqiang and Carroll, Joseph P and Wang, Shuqiu and Ran, Sheng and Broyles, Christopher and Siddiquee, Hasan and Butch, Nicholas P and Saha, Shanta R and Paglione, Johnpierre and Davis, JC S{\'e}amus and others},
  journal={Nature},
  volume={618},
  number={7967},
  pages={921--927},
  year={2023},
  publisher={Nature Publishing Group UK London}
}

@article{lee2006doping,
  title={Doping a Mott insulator: Physics of high-temperature superconductivity},
  author={Lee, Patrick A and Nagaosa, Naoto and Wen, Xiao-Gang},
  journal={Reviews of modern physics},
  volume={78},
  number={1},
  pages={17--85},
  year={2006},
  publisher={APS}
}

@article{paramekanti2001projected,
  title={Projected wave functions and high temperature superconductivity},
  author={Paramekanti, Arun and Randeria, Mohit and Trivedi, Nandini},
  journal={Physical review letters},
  volume={87},
  number={21},
  pages={217002},
  year={2001},
  publisher={APS}
}

@article{gros1989physics,
  title={Physics of projected wavefunctions},
  author={Gros, Claudius},
  journal={Annals of Physics},
  volume={189},
  number={1},
  pages={53--88},
  year={1989},
  publisher={Elsevier}
}

@article{gutzwiller1963effect,
  title={Effect of correlation on the ferromagnetism of transition metals},
  author={Gutzwiller, Martin C},
  journal={Physical Review Letters},
  volume={10},
  number={5},
  pages={159},
  year={1963},
  publisher={APS}
}

@article{yokoyama1987variational1,
  title={Variational monte-carlo studies of hubbard model. {I}},
  author={Yokoyama, Hisatoshi and Shiba, Hiroyuki},
  journal={Journal of the Physical Society of Japan},
  volume={56},
  number={4},
  pages={1490--1506},
  year={1987},
  publisher={The Physical Society of Japan}
}

@article{yokoyama1987variational2,
  title={Variational Monte-Carlo studies of hubbard model. {II}},
  author={Yokoyama, Hisatoshi and Shiba, Hiroyuki},
  journal={Journal of the Physical Society of Japan},
  volume={56},
  number={10},
  pages={3582--3592},
  year={1987},
  publisher={The Physical Society of Japan}
}

@article{himeda2002stripe,
  title={Stripe states with spatially oscillating d-wave superconductivity in the two-dimensional {$t-t'-J$} model},
  author={Himeda, A and Kato, T and Ogata, M},
  journal={Physical review letters},
  volume={88},
  number={11},
  pages={117001},
  year={2002},
  publisher={APS}
}

@article{motrunich2005variational,
  title={Variational study of triangular lattice spin-1/ 2 model with ring exchanges and spin liquid state in {$\kappa$-(ET)$_2$Cu$_2$(CN)$_3$}},
  author={Motrunich, Olexei I},
  journal={Physical Review B—Condensed Matter and Materials Physics},
  volume={72},
  number={4},
  pages={045105},
  year={2005},
  publisher={APS}
}

@article{iqbal2013gapless,
  title={Gapless spin-liquid phase in the kagome spin-1/2 Heisenberg antiferromagnet},
  author={Iqbal, Yasir and Becca, Federico and Sorella, Sandro and Poilblanc, Didier},
  journal={Physical Review B—Condensed Matter and Materials Physics},
  volume={87},
  number={6},
  pages={060405},
  year={2013},
  publisher={APS}
}

@article{edegger2007gutzwiller,
  title={Gutzwiller--RVB theory of high-temperature superconductivity: Results from renormalized mean-field theory and variational Monte Carlo calculations},
  author={Edegger, Bernhard and Muthukumar, Vangal N and Gros, Claudius},
  journal={Advances in Physics},
  volume={56},
  number={6},
  pages={927--1033},
  year={2007},
  publisher={Taylor \& Francis}
}

@article{lee2014amperean,
  title={Amperean pairing and the pseudogap phase of cuprate superconductors},
  author={Lee, Patrick A},
  journal={Physical Review X},
  volume={4},
  number={3},
  pages={031017},
  year={2014},
  publisher={APS}
}

@article{berg2009striped,
  title={Striped superconductors: how spin, charge and superconducting orders intertwine in the cuprates},
  author={Berg, Erez and Fradkin, Eduardo and Kivelson, Steven A and Tranquada, John M},
  journal={New Journal of Physics},
  volume={11},
  number={11},
  pages={115004},
  year={2009},
  publisher={IOP Publishing}
}

@article{arovas1984fractional,
  title={Fractional statistics and the quantum Hall effect},
  author={Arovas, Daniel and Schrieffer, John R and Wilczek, Frank},
  journal={Physical review letters},
  volume={53},
  number={7},
  pages={722},
  year={1984},
  publisher={APS}
}

@article{jain1989composite,
  title={Composite-fermion approach for the fractional quantum Hall effect},
  author={Jain, Jainendra K},
  journal={Physical review letters},
  volume={63},
  number={2},
  pages={199},
  year={1989},
  publisher={APS}
}

@misc{supp,
  title = {Please refer to the supplemental material for further details on additional models, settings, properties, and results for higher-charge superconductors, as well as our computational methods for their expectation values. }
}

@article{read2000paired,
  title={Paired states of fermions in two dimensions with breaking of parity and time-reversal symmetries and the fractional quantum Hall effect},
  author={Read, Nicholas and Green, Dmitry},
  journal={Physical Review B},
  volume={61},
  number={15},
  pages={10267},
  year={2000},
  publisher={APS}
}

@article{rice1995sr2ruo4,
  title={{Sr$_2$RuO$_4$}: an electronic analogue of 3He?},
  author={Rice, TM and Sigrist, M},
  journal={Journal of Physics: Condensed Matter},
  volume={7},
  number={47},
  pages={L643},
  year={1995},
  publisher={IOP Publishing}
}

@article{mackenzie2003superconductivity,
  title={The superconductivity of {Sr$_2$RuO$_4$} and the physics of spin-triplet pairing},
  author={Mackenzie, Andrew Peter and Maeno, Yoshiteru},
  journal={Reviews of Modern Physics},
  volume={75},
  number={2},
  pages={657},
  year={2003},
  publisher={APS}
}

@article{luke1998time,
  title={Time-reversal symmetry-breaking superconductivity in {Sr$_2$RuO$_4$}},
  author={Luke, G Ml and Fudamoto, Y and Kojima, KM and Larkin, MI and Merrin, J and Nachumi, B and Uemura, YJ and Maeno, Y and Mao, ZQ and Mori, Y and others},
  journal={Nature},
  volume={394},
  number={6693},
  pages={558--561},
  year={1998},
  publisher={Nature Publishing Group UK London}
}

@article{nelson2004odd,
  title={Odd-parity superconductivity in {Sr$_2$RuO$_4$}},
  author={Nelson, KD and Mao, ZQ and Maeno, Y and Liu, Ying},
  journal={Science},
  volume={306},
  number={5699},
  pages={1151--1154},
  year={2004},
  publisher={American Association for the Advancement of Science}
}

@article{kallin2016chiral,
  title={Chiral superconductors},
  author={Kallin, Catherine and Berlinsky, John},
  journal={Reports on Progress in Physics},
  volume={79},
  number={5},
  pages={054502},
  year={2016},
  publisher={IOP Publishing}
}

@article{zhang2011entanglement,
  title={Entanglement entropy of critical spin liquids},
  author={Zhang, Yi and Grover, Tarun and Vishwanath, Ashvin},
  journal={Physical review letters},
  volume={107},
  number={6},
  pages={067202},
  year={2011},
  publisher={APS}
}

@article{zhou2017quantum,
  title={Quantum spin liquid states},
  author={Zhou, Yi and Kanoda, Kazushi and Ng, Tai-Kai},
  journal={Reviews of Modern Physics},
  volume={89},
  number={2},
  pages={025003},
  year={2017},
  publisher={APS}
}

@article{iqbal2011projected,
  title={Projected wave function study of {Z2} spin liquids on the kagome lattice for the spin-1/2 quantum Heisenberg antiferromagnet},
  author={Iqbal, Yasir and Becca, Federico and Poilblanc, Didier},
  journal={Physical Review B—Condensed Matter and Materials Physics},
  volume={84},
  number={2},
  pages={020407},
  year={2011},
  publisher={APS}
}

@article{geshkenbein1987vortices,
  title={Vortices with half magnetic flux quanta in ‘‘heavy-fermion’’superconductors},
  author={Geshkenbein, Vadim B and Larkin, Anatoly I and Barone, Antonio},
  journal={Physical Review B},
  volume={36},
  number={1},
  pages={235},
  year={1987},
  publisher={APS}
}

@article{kee2000half,
  title={Half-quantum vortex and d-soliton in {Sr$_2$RuO$_4$}},
  author={Kee, Hae-Young and Kim, Yong Baek and Maki, Kazumi},
  journal={arXiv preprint cond-mat/0005510},
  year={2000}
}

@article{sigrist1989low,
  title={Low-field magnetic response of complex superconductors},
  author={Sigrist, M and Rice, TM and Ueda, K},
  journal={Physical review letters},
  volume={63},
  number={16},
  pages={1727},
  year={1989},
  publisher={APS}
}

@article{rampp2022integer,
  title={Integer and fractionalized vortex lattices and off-diagonal long-range order},
  author={Rampp, Michael A and Schmalian, J{\"o}rg},
  journal={Journal of Physics Communications},
  volume={6},
  number={5},
  pages={055013},
  year={2022},
  publisher={IOP Publishing}
}

@article{LI20242328,
title = {Charge {$4e$} superconductor: A wavefunction approach},
journal = {Science Bulletin},
volume = {69},
number = {15},
pages = {2328-2331},
year = {2024},
issn = {2095-9273},
doi = {https://doi.org/10.1016/j.scib.2024.06.002},
url = {https://www.sciencedirect.com/science/article/pii/S2095927324003980},
author = {Pengfei Li and Kun Jiang and Jiangping Hu}
}

@article{zhang2024higgs,
  title={Higgs-Leggett mechanism for the elusive {$\phi_0/3= hc/6e$} oscillation in Little-Parks setup of Kagome superconductor {CsV$_3$Sb$_5$}},
  author={Zhang, Ling-Feng and Wang, Zhi and Hu, Xiao},
  journal={Communications Physics},
  volume={7},
  number={1},
  pages={210},
  year={2024},
  publisher={Nature Publishing Group UK London}
}

@article{song2025phase,
  title={Phase coherence of charge-6e superconductors with a frustrated Kagome XY antiferromagnet},
  author={Song, Feng-Feng and Zhang, Guang-Ming},
  journal={Chinese Physics Letters},
  volume={42},
  number={3},
  pages={037401},
  year={2025},
  publisher={IOP Publishing}
}

@article{varma2023extended,
  title={Extended superconducting fluctuation region and {$6e$} and {$4e$} flux quantization in a kagome compound with a normal state of {3Q} order},
  author={Varma, Chandra M and Wang, Ziqiang},
  journal={Physical Review B},
  volume={108},
  number={21},
  pages={214516},
  year={2023},
  publisher={APS}
}

@article{han2022understanding,
  title={Understanding resistance oscillation in the {CsV$_3$Sb$_5$} superconductor},
  author={Han, Jung Hoon and Lee, Patrick A},
  journal={Physical Review B},
  volume={106},
  number={18},
  pages={184515},
  year={2022},
  publisher={APS}
}

@article{heitmann2019combined,
  title={Combined use of translational and spin-rotational invariance for spin systems},
  author={Heitmann, Tjark and Schnack, J{\"u}rgen},
  journal={Physical Review B},
  volume={99},
  number={13},
  pages={134405},
  year={2019},
  publisher={APS}
}

@article{mizusaki2004quantum,
  title={Quantum-number projection in the path-integral renormalization group method},
  author={Mizusaki, Takahiro and Imada, Masatoshi},
  journal={Physical Review B},
  volume={69},
  number={12},
  pages={125110},
  year={2004},
  publisher={APS}
}

@article{jimenez2012projected,
  title={Projected hartree--fock theory},
  author={Jim{\'e}nez-Hoyos, Carlos A and Henderson, Thomas M and Tsuchimochi, Takashi and Scuseria, Gustavo E},
  journal={The Journal of chemical physics},
  volume={136},
  number={16},
  year={2012},
  publisher={AIP Publishing}
}

@article{nikvsic2006beyond,
  title={Beyond the relativistic mean-field approximation. {II}. Configuration mixing of mean-field wave functions projected on angular momentum and particle number},
  author={Nik{\v{s}}i{\'c}, Tamara and Vretenar, Dario and Ring, Peter},
  journal={Physical Review C—Nuclear Physics},
  volume={74},
  number={6},
  pages={064309},
  year={2006},
  publisher={APS}
}

@article{sogo2009critical,
  title={Critical temperature for $alpha $-particle condensation within a momentum-projected mean-field approach},
  author={Sogo, T and Lazauskas, R and R{\"o}pke, G and Schuck, P},
  journal={Physical Review C},
  volume={79},
  pages={051301},
  year={2009}
}

@article{mermin1979topological,
  title={The topological theory of defects in ordered media},
  author={Mermin, N David},
  journal={Reviews of Modern Physics},
  volume={51},
  number={3},
  pages={591},
  year={1979},
  publisher={APS}
}

@article{zhao2021cascade,
  title={Cascade of correlated electron states in the kagome superconductor {CsV$_3$Sb$_5$}},
  author={Zhao, He and Li, Hong and Ortiz, Brenden R and Teicher, Samuel ML and Park, Takamori and Ye, Mengxing and Wang, Ziqiang and Balents, Leon and Wilson, Stephen D and Zeljkovic, Ilija},
  journal={Nature},
  volume={599},
  number={7884},
  pages={216--221},
  year={2021},
  publisher={Nature Publishing Group UK London}
}

@article{zhang2012quasiparticle,
  title={Quasiparticle statistics and braiding from ground-state entanglement},
  author={Zhang, Yi and Grover, Tarun and Turner, Ari and Oshikawa, Masaki and Vishwanath, Ashvin},
  journal={Physical Review B—Condensed Matter and Materials Physics},
  volume={85},
  number={23},
  pages={235151},
  year={2012},
  publisher={APS}
}

@article{grover2013entanglement,
  title={Entanglement entropy as a portal to the physics of quantum spin liquids},
  author={Grover, Tarun and Zhang, Yi and Vishwanath, Ashvin},
  journal={New Journal of Physics},
  volume={15},
  number={2},
  pages={025002},
  year={2013},
  publisher={IOP Publishing}
}

@article{zhang2013establishing,
  title={Establishing non-Abelian topological order in Gutzwiller-projected Chern insulators via entanglement entropy and modular S-matrix},
  author={Zhang, Yi and Vishwanath, Ashvin},
  journal={Physical Review B—Condensed Matter and Materials Physics},
  volume={87},
  number={16},
  pages={161113},
  year={2013},
  publisher={APS}
}

@article{gruner1988dynamics,
  title={The dynamics of charge-density waves},
  author={Gr{\"u}ner, George},
  journal={Reviews of modern physics},
  volume={60},
  number={4},
  pages={1129},
  year={1988},
  publisher={APS}
}

@article{gruner1994dynamics,
  title={The dynamics of spin-density waves},
  author={Gr{\"u}ner, G},
  journal={Reviews of modern physics},
  volume={66},
  number={1},
  pages={1},
  year={1994},
  publisher={APS}
}

@article{wang2014charge,
  title={Charge-density-wave order with momentum (2{$Q$}, 0) and (0, 2{$Q$}) within the spin-fermion model: Continuous and discrete symmetry breaking, preemptive composite order, and relation to pseudogap in hole-doped cuprates},
  author={Wang, Yuxuan and Chubukov, Andrey},
  journal={Physical Review B},
  volume={90},
  number={3},
  pages={035149},
  year={2014},
  publisher={APS}
}

@article{keimer2015quantum,
  title={From quantum matter to high-temperature superconductivity in copper oxides},
  author={Keimer, Bernhard and Kivelson, Steven A and Norman, Michael R and Uchida, Shinichi and Zaanen, J},
  journal={Nature},
  volume={518},
  number={7538},
  pages={179--186},
  year={2015},
  publisher={Nature Publishing Group UK London}
}

@article{fernandes2019intertwined,
  title={Intertwined vestigial order in quantum materials: Nematicity and beyond},
  author={Fernandes, Rafael M and Orth, Peter P and Schmalian, J{\"o}rg},
  journal={Annual Review of Condensed Matter Physics},
  volume={10},
  number={1},
  pages={133--154},
  year={2019},
  publisher={Annual Reviews}
}

@article{davis2013concepts,
  title={Concepts relating magnetic interactions, intertwined electronic orders, and strongly correlated superconductivity},
  author={Davis, JC S{\'e}amus and Lee, Dung-Hai},
  journal={Proceedings of the National Academy of Sciences},
  volume={110},
  number={44},
  pages={17623--17630},
  year={2013},
  publisher={National Academy of Sciences}
}

@article{gao2025topological,
  title={Topological Charge-2ne Superconductors},
  author={Gao, Zhi-Qiang and Wang, Yan-Qi and Yang, Hui and Wu, Congjun},
  journal={arXiv preprint arXiv:2512.21325},
  year={2025}
}

@article{shi2025non,
  title={Non-Abelian topological superconductivity from melting Abelian fractional Chern insulators},
  author={Shi, Zhengyan Darius and Senthil, T},
  journal={arXiv preprint arXiv:2512.17996},
  year={2025}
}

@article{yerin2023multiple,
  title={Multiple-q current states in a multicomponent superconducting channel},
  author={Yerin, Yuriy and Drechsler, Stefan-Ludwig and Cuoco, Mario and Petrillo, Caterina},
  journal={Journal of Physics: Condensed Matter},
  volume={35},
  number={50},
  pages={505601},
  year={2023},
  publisher={IOP Publishing}
}

@article{volovik2024fermionic,
  title={Fermionic quartet and vestigial gravity},
  author={Volovik, Grigorii Efimovich},
  journal={JETP Letters},
  volume={119},
  number={4},
  pages={330--334},
  year={2024},
  publisher={Springer}
}

@article{saminadayar1997observation,
  title={Observation of the e/3 fractionally charged Laughlin quasiparticle},
  author={Saminadayar, L and Glattli, DC and Jin, Y and Etienne, B c-m},
  journal={Physical Review Letters},
  volume={79},
  number={13},
  pages={2526},
  year={1997},
  publisher={APS}
}

@article{tsui1982two,
  title={Two-dimensional magnetotransport in the extreme quantum limit},
  author={Tsui, Daniel C and Stormer, Horst L and Gossard, Arthur C},
  journal={Physical Review Letters},
  volume={48},
  number={22},
  pages={1559},
  year={1982},
  publisher={APS}
}

@article{yamashita2008thermodynamic,
  title={Thermodynamic properties of a spin-1/2 spin-liquid state in a $\kappa$-type organic salt},
  author={Yamashita, Satoshi and Nakazawa, Yasuhiro and Oguni, Masaharu and Oshima, Yugo and Nojiri, Hiroyuki and Shimizu, Yasuhiro and Miyagawa, Kazuya and Kanoda, Kazushi},
  journal={Nature Physics},
  volume={4},
  number={6},
  pages={459--462},
  year={2008},
  publisher={Nature Publishing Group UK London}
}

@article{yamashita2010highly,
  title={Highly mobile gapless excitations in a two-dimensional candidate quantum spin liquid},
  author={Yamashita, Minoru and Nakata, Norihito and Senshu, Yoshinori and Nagata, Masaki and Yamamoto, Hiroshi M and Kato, Reizo and Shibauchi, Takasada and Matsuda, Yuji},
  journal={Science},
  volume={328},
  number={5983},
  pages={1246--1248},
  year={2010},
  publisher={American Association for the Advancement of Science}
}

\end{document}